\documentclass[journal]{IEEEtran}
\newcommand{\PreserveBackslash}[1]{\let\temp=\\#1\let\\=\temp}

\usepackage{color}
\usepackage{array}
\usepackage{multirow}
\usepackage{textcomp}
\usepackage{xcolor}
\usepackage{cite}
\usepackage{graphicx}
\usepackage{cite}
\usepackage{subfigure}
\usepackage[T1]{fontenc}

\graphicspath{{./fig/}}

\begin{document}

\title{Novel Digital Camera with the PCIe Interface}

\author{
	D.~Makowski~\IEEEmembership{Senior Member,~IEEE}, A.~Mielczarek	
	\thanks{D.~Makowski, A.~Mielczarek are with the Lodz University of Technology, Poland (e-mail: dmakow@dmcs.pl)}
}

\maketitle

\thispagestyle{empty}
\begin{abstract}
Digital cameras are commonly used for diagnostic purposes in large-scale physics experiments.
A typical image diagnostic system consists of an optical setup, digital camera, frame grabber, image processing CPU, and data analysis tool.
The standard architecture of the imaging system has a number of disadvantages. Data transmitted from a camera are buffered multiple times and must be converted between various protocols before they are finally transmitted to the host memory. Such an architecture makes the system quite complicated, limits its performance and, in consequence, increases its price. The limitations are even more critical for control or protection systems operating in real-time.

Modern megapixel cameras generate large data throughput, easily exceeding 10~Gb/s, which often requires some additional processing on the host side.
The optimal system architecture should assure low overhead and high performance of the data transmission and processing. It is particularly important during the processing of data streams from several imaging devices, which can be as high as several terabits per second.

A novel architecture of image acquisition and processing system based on the PCI Express interface was proposed to meet the requirements of real-time imaging systems applied in large-scale physics experiments. The architecture allows to transfer an image stream directly from the camera to the data processing unit and therefore significantly decreases the overhead and improves performance.

Two various architectures will be presented, compared and discussed in the paper.


\end{abstract}
\begin{IEEEkeywords}
Digital Camera, Image Acquisition, Image Processing, Camera Interface, PCI Express Interface, Programmable Device, Plasma Diagnostics, Beam Diagnostics  
\end{IEEEkeywords}

\section{Introduction}

\IEEEPARstart{D}{igital} cameras are usually applied  for diagnostic purposes in various large-scale physics experiments. A good example is the imaging system for beam diagnostics in particle accelerators that allow measuring the beam charge profile or its transverse emittance~\cite{tang2010application}.
Another example is plasma diagnostics in tokamaks~\cite{balboa2012upgrade} or stellarators~\cite{konig2010diag_short}, such as the imaging systems of the ITER tokamak that require more than 200 digital cameras working in the visible, infrared, and gamma radiation range~\cite{simrock2013diagnostic}. Tokamak vision systems should allow for the image acquisition and processing with the resolution of 1 to 8 million pixels registered within the range of 50 to 50000 frames per second. For example, a digital 8-bit camera with a resolution of 1 megapixel working with a frame rate of 1000 frames per second generates a stream of data not less than 8~Gb/s. In this situation, the tokamak image acquisition system, which provides a stream of data from 10 cameras, requires a data throughput of at least 80~Gb/s.

\begin{figure}[htb]
\centering
\includegraphics[width=0.5\textwidth]{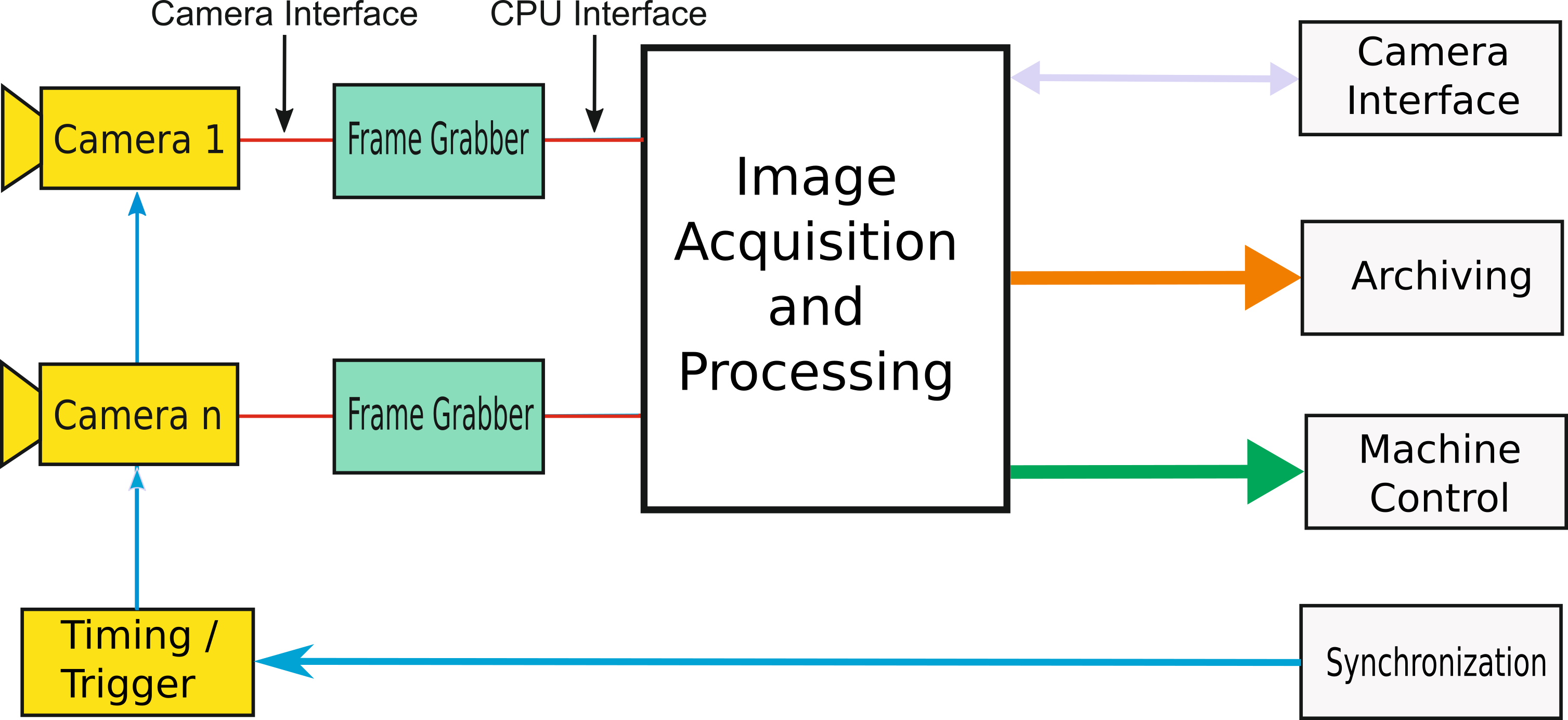}
\caption{A typical Image Acquisition System used in large-scale physical experiments}
\label{fig:IAS}
\end{figure}

A classic approach to capturing and computing video data is presented in Fig.~\ref{fig:IAS}. The system interfaces with several cameras, performs data processing and provides the resulting streams to archiving and machine control networks. The system should provide synchronisation and timestamping of the images from the cameras with an accuracy better than 50 ns(rms). The image acquisition system needs to process the images in real-time, providing information to the safety subsystems in order to protect the tokamak against damage. Such data should be processed in less than 100~\textmu{}s. Moreover, the data used to control the plasma should be delivered in less than tens of ms.

Field Programmable Gate Array (FPGA) devices are applied for low latency and relatively simple image processing to meet real-time constraints of several ms or less. However, more complex, but less time-critical algorithms could be implemented using a standard CPU with a parallel GPU (Graphics Processing Unit) acceleration. 

The development of an efficient and simple architecture suitable for processing of large amounts of data in real-time is a challenging task especially in systems computing data from cameras equipped with various camera interfaces. Unification and standardisation of camera interface could simplify the architecture, however the system still suffer from multiple data copying and buffering.


\section{Architecture}

\begin{figure}[htb]
\centering
\includegraphics[width=0.5\textwidth]{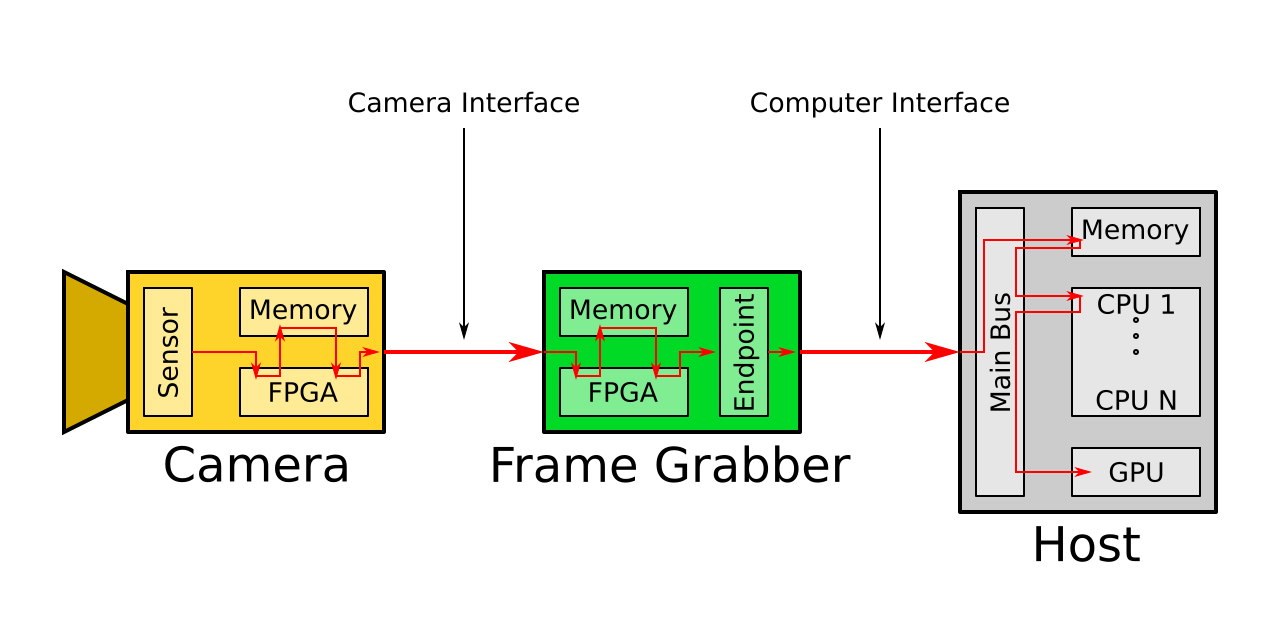}
\caption{A classical approach to design image acquisition system}
\label{fig:old_ias}
\end{figure}

The standard architecture of imaging system (Fig.~\ref{fig:old_ias}) consists of:
\begin{itemize}
\item a Camera -- converting optical signal to electrical and transferring it via a dedicated Camera Interface (CI) protocol,
\item a Frame Grabber -- capturing data from the Camera Interface and converting to the PCI Express (PCIe) host interface,
\item a Host Computer -- receiving and processing raw data stream, controlling the camera.
\end{itemize}

Camera sensor registers images that are further processed with either a dedicated Application-Specific Integrated Circuit (ASIC) or more commonly with a Field-Programmable Gate Array. Buffered data are transmitted using a dedicated camera interface. The camera is often providing simple image pre-processing such as: gain, offset calibration or image corrections, however it cannot execute custom algorithms provided by the user. The video stream is then encoded to one of several common protocols, that can be used by the Camera Interface. Various camera interfaces were standardised including, among others: IEEE-1394, Camera Link (CL), Camera Link High Speed (CLHS), GigE Vision, CoaXPress (CXP), and Universal Serial Bus (USB). Depending on the chosen interface standard this link can have a maximum length limit from several up to several hundreds of meters.

The data arriving through the Camera Interface cannot be directly provided to the processing unit (CPU or GPU). The CI protocols are oriented on data streams whereas the host bus architectures are based on the memory map concept, where each chunk of data has an assigned address. A dedicated Frame Grabber device is required to buffer images and convert data from the CI to the PCIe interface.  
The frame grabber card is usually implemented with a programmable device (FPGA) with the manufacturer firmware and cannot be used for image processing. Real algorithms are computed in CPU or GPU. This requires further data buffering and copying. 

The main drawbacks of the standard architecture are multiple data buffering, copying and translating to various interfaces that significantly complicate the design, reduce reliability, increase the total latency and the real-time performance. There is also waste of processing power in the image acquisition chain. The available FPGA devices in the camera and frame grabber module cannot be easily used for image processing.

We propose to modify the architecture of the imaging system in such a way that the camera could send data directly to the host memory. The PCI Express interface should be moved as close to the camera sensor as possible for an efficient data transmission with low buffering overhead. In such an application the PCI Express endpoint could be implemented directly in the camera FPGA, and therefore devices will be visible in the PCIe space. Image data could be transferred directly from the camera to CPU and/or GPU using efficient DMA (Direct Memory Access) transfers. The PCIe Camera Interface could be also implemented using an optical fibre that assures galvanic isolation and reliable transmission up too 300~m according to the authors' experience.  

This approach eliminates the need for a separate Camera Interface and for a Frame Grabber. The Camera Interface is often the main bottleneck of the whole path, whereas the Frame Grabber is a source of an additional latency. The FPGA and memory available in the camera could be used for the PCIe transmission and therefore there is no additional costs for such a design.   

\begin{figure}[htb]
\centering
\includegraphics[width=0.5\textwidth]{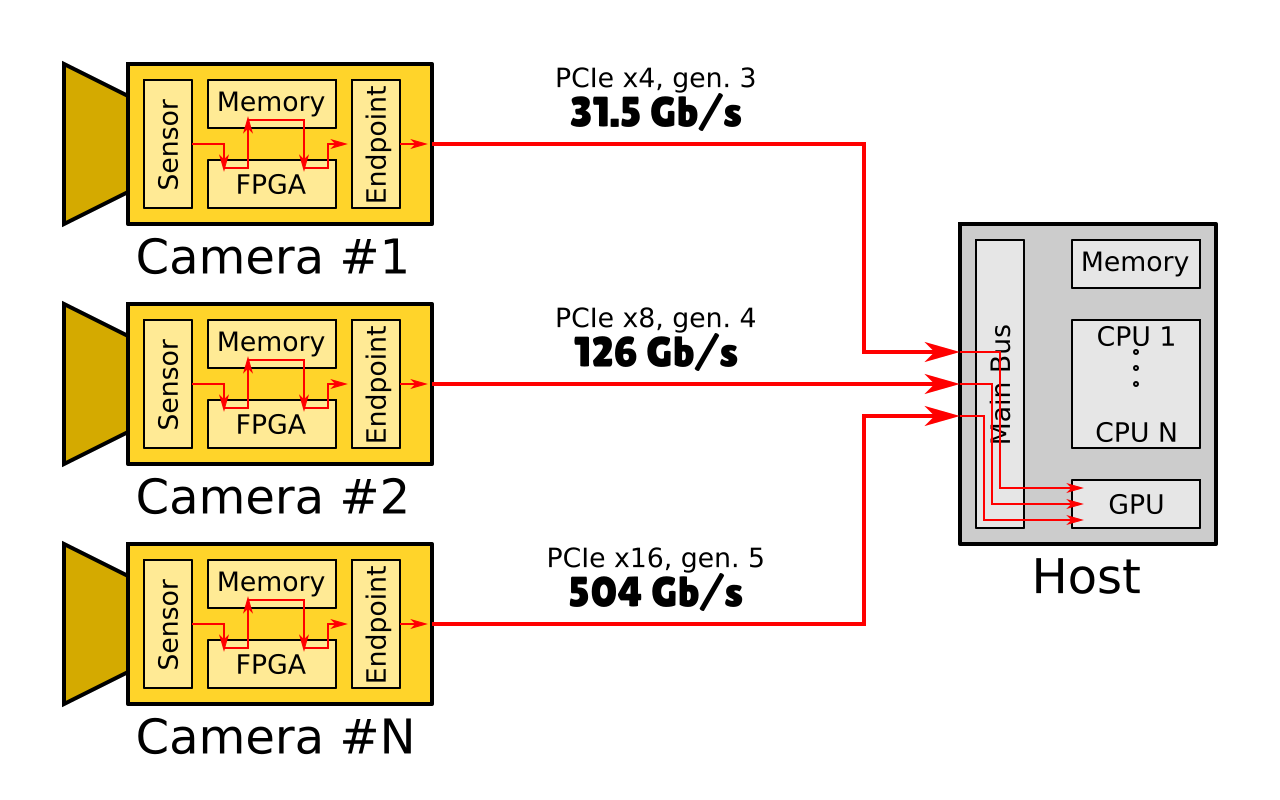}
\caption{The image Acquisition System based on the PCI Express interface}
\label{fig:new_ias}
\end{figure}

The architecture of the novel image acquisition system proposed by the authors is shown in Fig.~\ref{fig:new_ias}.

A direct application of the PCIe interface in the camera enables unprecedented transmission rates. A single (x1) PCIe gen.~3 lane offers more throughput (\textasciitilde{}7.88~Gb/s) than the highest speed mode of the legacy Camera Link interface (\textasciitilde{}7.14~Gb/s). The two further generations of the standard allow to transfer with data rates up to \textasciitilde{}16~Gb/s (PCIe gen. 4) and \textasciitilde{}32~Gb/s (gen. 5). The interface throughput could be also easily scaled and adjusted for camera needs reaching up too 504~Gb/s when 16 lanes of PCIe gen. 3 are used, see Fig.~\ref{fig:new_ias}.

The camera architecture based on the PCI Express standard allows to obtain the most optimal design for the image acquisition and processing system. This allows to decrease the total latency, decrease the price of the system and improve the total performance. 
Since the PCIe is a well standardised interface and it is maintained by the Peripheral Component Interconnect Special Interest Group (PCI-SIG), it guarantees an easy and simple upgrade path to the future revisions and full compatibility with future computers.

\bibliographystyle{IEEEtran}
\bibliography{./bib/rt18,./bib/rt20}

\end{document}